\documentstyle[11pt,amssymb]{article}

\def\dalemb#1#2{{\vbox{\hrule height .#2pt
        \hbox{\vrule width.#2pt height#1pt \kern#1pt
                \vrule width.#2pt}
        \hrule height.#2pt}}}

\def\cF{{\cal F}}
\def\cA{{\cal A}}

\def\0{{\sst{(0)}}}
\def\1{{\sst{(1)}}}
\def\2{{\sst{(2)}}}
\def\3{{\sst{(3)}}}
\def\4{{\sst{(4)}}}
\def\5{{\sst{(5)}}}
\def\6{{\sst{(6)}}}
\def\7{{\sst{(7)}}}
\def\8{{\sst{(8)}}}

\def\Z{\rlap{\sf Z}\mkern3mu{\sf Z}}
\def\R{\rlap{\rm I}\mkern3mu{\rm R}}

\def\ep{\epsilon}
\def\td{\tilde}
\def\wtd{\widetilde}

\let\a=\alpha    
    \let\k=\kappa

\def\nn{\nonumber} \def\bd{\begin{document}} \def\ed{\end{document}}
\def\ds{\documentstyle} \let\fr=\frac \let\bl=\bigl \let\br=\bigr
\let\Br=\Bigr \let\Bl=\Bigl 
\let\bm=\bibitem
\let\na=\nabla
\let\pa=\partial \let\ov=\overline 
\newcommand{\be}{\begin{equation}} 
\newcommand{\ee}{\end{equation}} 
\def\ba{\begin{array}}
\def\ea{\end{array}}
\def\ft#1#2{{\textstyle{{\scriptstyle #1}\over {\scriptstyle #2}}}}
\def\fft#1#2{{#1 \over #2}}
\def\del{\partial}
\def\sst#1{{\scriptscriptstyle #1}}
\def\oneone{\rlap 1\mkern4mu{\rm l}}
\def\ie{{\it i.e.\ }}
\def\via{{\it via}}
\def\semi{{\ltimes}}
\def\str{{\rm str}}
\def\jm{{\rm j}}
\def\im{{\rm i}}
\def\bOmega{{{\bar\Omega}}}
\def\Qn{{{Q_{\sst{\rm N}}}}}
\def\tX{{{\wtd X}}}

\def\mapright#1{\smash{\mathop{-\!\!\!-\!\!\!-\!\!\!-\!\!\!-\!\!\!
             \longrightarrow}\limits^{#1}}}
\def\maprightt#1#2{\smash{\mathop{-\!\!\!-\!\!\!-\!\!\!-\!\!\!-\!\!\!
             \longrightarrow}\limits^{#1}_{#2}}}

\newcommand{\ho}[1]{$\, ^{#1}$}
\newcommand{\hoch}[1]{$\, ^{#1}$}
\newcommand{\bea}{\begin{eqnarray}} 
\newcommand{\eea}{\end{eqnarray}} 
\newcommand{\ra}{\rightarrow}
\newcommand{\lra}{\longrightarrow}
\newcommand{\Lra}{\Leftrightarrow}
\newcommand{\ap}{\alpha^\prime}
\newcommand{\bp}{\tilde \beta^\prime}
\newcommand{\tr}{{\rm tr} }
\newcommand{\Tr}{{\rm Tr} }

\newcommand{\NP}{Nucl. Phys. }
\newcommand{\tamphys}{\it Center for Theoretical Physics\\
Texas A\&M University, College Station, Texas 77843}
\newcommand{\ens}{\it Laboratoire de Physique Th\'eorique de l'\'Ecole
Normale Sup\'erieure\hoch{2,3}\\
24 Rue Lhomond - 75231 Paris CEDEX 05}
\newcommand{\upenn}{\it Department of Physics and Astronomy\\
University of Pennsylvania, Philadelphia, Pennsylvania 19104}

\newcommand{\auth}{M. Cveti\v{c}\hoch{\dagger1}, 
H. L\"u\hoch{\dagger1} and C.N. Pope\hoch{\ddagger2}}

\begin{document}
\begin{flushright}
\hfill{CTP TAMU-21/00}\\
\hfill{UPR-895-T}\\
\hfill{hep-th/0007109}\\
\hfill{July, 2000}\\
\end{flushright}


\begin{center}
{ \large {\bf Consistent Warped-Space Kaluza-Klein Reductions,
Half-Maximal Gauged Supergravities and $CP^n$ Constructions}}

\vspace{15pt}
\auth

\vspace{15pt}

{\hoch{\dagger}\upenn}

\vspace{15pt}
{\hoch{\ddagger}\tamphys}

\vspace{40pt}

\underline{ABSTRACT}
\end{center}

   We obtain new consistent Kaluza-Klein embeddings of the gauged
supergravities with half of maximal supersymmetry in dimensions $D=7$,
6, 5 and 4.  They take the form of warped embeddings in type IIA, type
IIB, M-theory and type IIB respectively, and are obtained by
performing Kaluza-Klein circle reductions or T-duality transformations
on Hopf fibres in $S^3$ submanifolds of the previously-known sphere
reductions.  The new internal spaces are in some sense ``mirror
manifolds'' that are dual to the original internal spheres.  The
vacuum AdS solutions of the gauged supergravities then give rise to
warped products with these internal spaces.  As well as these
embeddings, which have singularities, we also construct new
non-singular warped Kaluza-Klein embeddings for the $D=5$ and $D=4$
gauged supergravities.  The geometry of the internal spaces in these
cases leads us to study Fubini-Study metrics on complex projective
spaces in some detail.

{\vfill\leftline{}\vfill
\footnoterule
{\footnotesize \hoch{1} Research supported in part by DOE grant 
DE-FG02-95ER40893 \vskip -12pt} \vskip 14pt
{\footnotesize  \hoch{2} Research supported in part by DOE 
grant DE-FG03-95ER40917.\vskip  -12pt}}

\pagebreak
\setcounter{page}{1}

\section{Introduction}

   It has long been known that gravity coupled to antisymmetric
tensors can allow AdS$\times$Sphere solutions \cite{freund}.  Such
configurations occur for eleven-dimensional supergravity and type IIB
supergravity, where they give rise to solutions AdS$_4\times S^7$,
AdS$_7\times S^4$ and AdS$_5\times S^5$, which preserve maximal
supersymmetry \cite{kk}.  At the linearised level the fluctuations
around these backgrounds correspond to the fields of the associated
maximal gauged supergravities \cite{dupo,PTV,GM,KRV}.  It was expected
that the reductions on $S^7$, $S^4$ and $S^5$ would in fact be
consistent at the full non-linear level, and indeed for the $S^7$
reduction \cite{dewnic} and the $S^4$ reduction \cite{vann2} this has
been demonstrated.  As well as these maximally-supersymmetric
reductions, the explicit non-linear reductions with half of maximal
supersymmetry have been obtained for the $S^7$ and $S^4$ cases
\cite{d4gauge,d7gauge}, and in addition for the $S^5$ example
\cite{d5gauge}.  Although no results for the complete
maximally-supersymmetric reduction of type IIB supergravity on $S^5$
exist, further supporting evidence has been obtained by constructing
the complete $S^5$ reduction of the $SL(2,\R)$-singlet sector of the
type IIB theory \cite{clpst}.

    The higher-dimensional interpretation of six-dimensional gauged
supergravity is more subtle.  First of all, gauged supergravity in
$D=6$ can at most have half of the supersymmetry that is possible in
the ungauged theory \cite{romans1}.  It was suggested that the gauged
theory might be related to the massive type IIA supergravity
\cite{ferrara}.  Indeed, it was shown in \cite{oz} that the massive
type IIA theory admits a warped product of AdS$_6$ and $S^4$, with a
warp factor in front of the AdS$_6$ metric that depends on a
coordinate of the $S^4$.  This solution can be derived \cite{oz} as a
near-horizon limit of a semi-localised D4/D8-brane intersection
\cite{youm}.  The fully non-linear consistent embedding of the
six-dimensional $N=2$ $SU(2)$-gauged supergravity in the massive type
IIA theory was subsequently constructed in \cite{d6gauge}.  It was recently
observed \cite{clpv} that AdS$_6$ could also be embedded in type IIB
theory, as the near-horizon limit of a semi-localised intersecting
D3/D5/NS5-brane system. In this paper, we shall obtain the non-linear
embedding of the six-dimensional $N=2$ gauged supergravity in type IIB.

   Recently, it was shown that AdS$_5$ can also arise in a warped
spacetime solution of M-theory, as the near-horizon geometry
\cite{preoz2,oz2} of a semi-localised M5/M5-brane intersection \cite{youm}.
A large class of analogous solutions involving warped products of AdS
and an internal space were subsequently constructed \cite{clpv},
arising as semi-localised intersections of two or more $p$-branes.  In
all these cases, the warp factors that multiply the AdS metrics depend
only on certain coordinates of the internal spaces.  In this paper, we
shall consider those warped configurations that are associated with
intersecting branes that preserve half of maximal supersymmetry in
their near-horizon regions.

    As in the case of the maximally-supersymmetric direct-product
AdS$\times$Sphere solutions, one might expect that the occurrence of
the half-maximally supersymmetric warped AdS solutions would also
presage the possibility of obtaining the associated half-maximally
supersymmetric gauged supergravities by consistent Kaluza-Klein
reduction on the corresponding internal spaces.  

    In this paper we construct a variety of examples of such
half-maximally supersymmetric gauged supergravities, arising as
consistent Kaluza-Klein reductions.  We obtain them by starting with
the previously-known maximally-supersymmetric sphere reductions, and
exploiting the fact that the internal sphere $S^n$ can itself be
viewed as a foliation of $S^p\times S^q$ with $n=p+q+1$.  In all the
examples $n\ge 4$, and so we can arrange that at least one of $p$ or
$q$ is equal to 3.  We then perform a standard $S^1$ Kaluza-Klein
reduction on the $U(1)$ fibre of $S^3$ viewed as the Hopf bundle over
$S^2$.  The gauge groups that are compatible with the $S^p\times S^q$
structures are smaller than the $SO(n+1)$ groups of the
maximally-supersymmetric theories, and in fact in all cases the
subgroups that {\it are} compatible are precisely those of the
half-maximally supersymmetric supergravities.  These are the
theories that whose consistent reductions were obtained in
\cite{d7gauge,d6gauge,d5gauge,d4gauge}.

    In section 2 we carry out this procedure for the $N=2$
$SU(2)$-gauged theory in $D=7$, obtained as an $S^4$ reduction from
$D=11$.  The $S^4$ is viewed as a foliation of $S^3$ surfaces, on
which the $SU(2)$ gauge group acts transitively.  We perform a
reduction on the $U(1)$ Hopf fibres of $S^3$, thereby arriving at a
reduction of type IIA supergravity that yields the same $N=2$ gauged
theory in $D=7$.  In section 3 we consider the warped $S^4$ reduction
of the massive type IIA theory.  Since this is already half-maximally
supersymmetric, it is already compatible with the $S^3$ structure of
the foliation.  We reduce this on the Hopf fibres to $D=9$, and after
performing a T-duality transformation we arrive at a type IIB
embedding of the six-dimensional $N=2$ $SU(2)$-gauged supergravity.
In section 4 we start from the half-maximally supersymmetric reduction
of type IIB supergravity on $S^5$, whose $SU(2)\times U(1)$ gauge
group is precisely compatible with the $S^3\times S^1$ foliation of
$S^5$.  After a Hopf reduction and T-duality transformation, we obtain
a type IIA embedding of the five-dimensional $N=4$ $SU(2)\times
U(1)$-gauged supergravity.  In section 5 we start from the
half-maximally supersymmetric reduction of eleven-dimensional
supergravity on $S^7$, whose $SO(4)\sim SU(2)\times SU(2)$ gauge
group is compatible with an $S^3\times S^3$ foliation.  Here there are
two bundles, associated with the two $S^3$ factors, and so we are
able to reduce first to a type IIA embedding, and then after a second
reduction we can obtain an embedding of the four-dimensional $N=4$
$SO(4)$-gauged supergravity in type IIB.  In all these examples, the
embeddings that we obtain are of the form of warped Kaluza-Klein
reductions.  Indeed, the AdS vacuum solutions of the half-maximally
supersymmetric lower-dimensional theories all lift to give the warped
products that were found in \cite{preoz2,oz2,clpv}.

   A characteristic feature of these warped Kaluza-Klein reductions is
that the warp-factor that multiplies the lower-dimensional spacetime
metric is singular, tending to zero at one end of the range of a
coordinate in the internal space.\footnote{Non-singular embedding of
AdS$_5$ in M-theory and AdS$_3$ in type IIB were recently constructed
in \cite{fs,mn} and \cite{mn} respectively.}  In the cases of those originating
from the $S^5$ and $S^7$ reductions, a slightly more general type of
Hopf reduction can be performed, which again involves a warped product
structure, but now with an entirely non-singular warp factor.  This is
possible because in these two cases the foliating surfaces are
actually products of {\it two} odd-dimensional spheres ($S^3\times
S^1$ and $S^3\times S^3$ respectively).  Thus we have two natural
$U(1)$ Killing directions in each case, associated with the Hopf
fibres over $S^3$ or from the $S^1$ factor.  For a single step of
reduction it is now possible to take a linear combination of the two
$U(1)$ Killing directions, and use this as the circle for the $S^1$
reduction.  Since the radii of the two odd-dimensional spheres in the
foliation never vanish simultaneously, this means that the radius of
the circle on which the Kaluza-Klein reduction is performed never
vanishes.  As a consequence, the resulting warped Kaluza-Klein
reduction is then entirely non-singular.  In fact, for the most
natural choice of linear combination of the Killing directions, it
turns out that the internal space after the $S^1$ reduction is a
complex projective space; $CP^2$ in the $S^5$ case, and $CP^3$ in the
$S^7$ case.  In fact, the Kaluza-Klein reductions that we obtain in
these cases correspond, in the vacuum, to the AdS$_5\times CP^2$ and
AdS$_4\times CP^3$ backgrounds that were considered in
\cite{ads5s5,nilpop,susywosusy}.  The non-singular warped Kaluza-Klein
reductions are discussed in section 6.

   Motivated by the occurrence of $CP^2$ and $CP^3$ in the
non-singular warped reductions, in an appendix we study some of the
related geometrical aspects of the Fubini-Study metrics on complex
projective spaces.  We obtain general constructions for $CP^{m+n+1}$ in
terms of a product of $CP^m$ and $CP^n$ spaces.  Applying this to the
case $m=n=1$ gives a very simple explicit construction of the
Fubini-Study metric on $CP^3$.  

      Finally we note that the subjects of sphere compactification and
semi-localised intersecting $p$-brane solutions were extensively discussed
in the literature, see additional references, {\it e.g.} 
\cite{deWitnicolaiwarner,wnb,KLT,cvgub,ten,FGPW,vann1,BS,BSI,dist,BBS,clpsa,%
kkred,s3s4,prt,cs,ks,cksz,sabnew,cs2,cvet1,cvet2,tseyt1,horo,callan,tseyt2,%
lpint,ity,fs1,loewy,gkmt,youm2,cherkis}

\section{$D=7$ $SU(2)$-gauged $N=2$ supergravity from type IIA}

   The embedding of seven-dimensional $N=2$ $SU(2)$-gauged
supergravity in $D=11$ was obtained in \cite{d7gauge}, in a framework
where the 4-sphere is described as a foliation of $S^3$ surfaces.  In this
section, we take this construction as our starting point, and then
perform a Kaluza-Klein reduction on the $U(1)$ fibres of the $S^3$
foliations, thereby obtaining an embedding of the seven-dimensional
theory in type IIA supergravity.

   From \cite{d7gauge}, we have the Kaluza-Klein reduction Ansatz for
the $S^4$ reduction from $D=11$, where we truncate to the bosonic
sector of $N=2$ gauged $SU(2)$ supergravity in $D=7$.  The metric
reduction is given by
\be
d\hat s_{11}^2 = \Delta^{1/3}\, ds_7^2 + 2 g^{-2}\, X^3\, \Delta^{1/3}
\, d\xi^2 + \ft12 g^{-2}\, \Delta^{-2/3}\, X^{-1}\, c^2\, \sum_i
h_i^2\,,\label{d11metred}
\ee
where 
\bea
&&\Delta = X\, c^2 + X^{-4}\, s^2 \,,\nn\\
&&c\equiv \cos\xi \,,\qquad s\equiv \sin\xi\,,\\
&&h_i \equiv \sigma_i -g\, A_\1^i\,.\nn
\eea
The left-invariant 1-forms of $SU(2)$ are given by
\bea
\sigma_1 &=& \cos\psi\, d\theta + \sin\psi\, \sin\theta\,
d\varphi\,,\nn\\
\sigma_2 &=& -\sin\psi\, d\theta + \cos\psi\, \sin\theta\,
d\varphi\,,\label{1forms}\\
\sigma_3 &=& d\psi + \cos\theta\, d\varphi\,.\nn
\eea
$X$ is the scalar field in the seven-dimensional $N=2$ gauged
supergravity, and $A_\1^i$ are the $SU(2)$ gauge fields.

     It is evident that $\del/\del\varphi$ is a Killing direction, and
so we can perform a Kaluza-Klein circle reduction on the $\varphi$
coordinate.  In order to do this, it is convenient to make the
following redefinitions:
\bea
z&=& \fft1{\sqrt2\, g}\, \varphi\,,\nn\\
A_\1^{ij} &=& \ep_{ijk}\, A_\1^k\,,\nn\\
\mu_1 &=& \sin\theta\, \sin\psi\,,\qquad \mu_2 = \sin\theta\, \cos\psi\,,
\qquad \mu_3 = \cos\theta\,.
\eea
We also define the gauge-covariant exterior derivative
\be
D\mu^i = d\mu^i + g\, A_\1^{ij}\, \mu^j\,.
\ee
Note that the coordinates $\mu^i$ satisfy
\be
\mu^i\, \mu^i = 1\,.
\ee

    It now follows that the 1-forms $h_i$ can be written as
\be
h_i = -\ep_{ijk}\, \mu^j\, D\mu^k + \sqrt2\, g\ \mu^i\, (dz +
\cA_\1)\,,
\ee
where
\be
\cA_\1 = \fft1{\sqrt2\, g}\, \cos\theta\, d\psi + \ft1{2\sqrt2}\,
\ep_{ijk}\, A_\1^{ij}\, \mu^k\,.\label{kkvector}
\ee
Of course the first term could be expressed in terms of an object
$\omega_\1$ such that $d\omega_\1=\Omega_\2$, the volume form of the
2-sphere, rather than using $\cos\theta\, d\psi$.   It also follows that
\be
\sum_i h_i^2 = \sum_i (D\mu^i)^2 + 2 g^2\, (dz + \cA_\1)^2\,,
\label{hexp0}
\ee

    Substituting this expression into (\ref{d11metred}), we obtain the
eleven-dimensional metric in the form
\bea
d\hat s_{11}^2 &=& \Delta^{1/3}\, ds_7^2 + 2 g^{-2}\, X^3\, \Delta^{1/3}
\, d\xi^2 + \ft12 g^{-2}\, \Delta^{-2/3}\, X^{-1}\, c^2\, \sum_i
(D\mu^i)^2\nn\\
&& + \Delta^{-2/3}\, X^{-1}\, c^2\, (dz+ \cA_\1)^2\,.\label{d11metred2}
\eea
Comparing this with the standard Kaluza-Klein $S^1$ reduction from $D=11$
to $D=10$,
\be
d\hat s_{11}^2 = e^{-\fft16\phi}\, ds_{10}^2 + e^{\fft43\phi}\, (dz
+ \cA_\1)^2\,,\label{d11d10red}
\ee
we find that the metric Ansatz describing the embedding of seven-dimensional
$N=2$ gauged $SU(2)$ supergravity in $D=11$ can be reinterpreted as an
embedding in type IIA supergravity, with the metric and dilaton given by
\bea
ds_{10}^2 &=& X^{-1/8}\, c^{1/4} \, \Big[ \Delta^{1/4}\, ds_7^2 + 
2g^{-2}\, X^3\, \Delta^{1/4}\, d\xi^2 + 
\ft12 g^{-2}\, X^{-1}\, \Delta^{-3/4}\, c^2\,
\sum_i(D\mu^i)^2\Big]\,,\nn\\
e^{\phi} &=& X^{-3/4}\, \Delta^{-1/2}\, c^{3/2}\,.
\eea
The Ansatz for the vector potential of the type IIA theory is given by
(\ref{kkvector}).  The metric of the vacuum solution corresponds to
taking $X=1$, $A_\1^{ij}=0$ and $ds_7^2$ to be AdS$_7$.  This solution
can be viewed as the near-horizon limit of the semi-localised
NS5/D6-brane system \cite{clpv}.
 
   We can now reduce the original $S^4$ Ansatz for the 3-form
potential $\hat A_\3$ of $D=11$ supergravity in an analogous way.  It
is given in \cite{d7gauge}, and takes the form
\be
\hat A_\3 = s\, A_\3 + \ft1{2\sqrt2}\, g^{-3}\, (2s + s\, c^2\,
X^{-4}\, \Delta^{-1} )\, \ep_\3 - \ft1{\sqrt2} g^{-2}\, s\,
F_\2^i\wedge h_i - \ft1{\sqrt2}\, g^{-1}\, s\, \omega_\3\,,
\ee
where 
\be
\ep_\3\equiv h_1\wedge h_2\wedge h_3 \,,
\ee
and
\be
\omega_\3 \equiv A_\1^i\wedge F_\2^i - \ft16 g\, \ep_{ijk}\,
A_\1^i\wedge A_\1^j \wedge A_\1^k\,.
\ee

    By comparing this with the standard $S^1$ reduction of $\hat
A_\3$ to $D=10$, with ten-dimensional fields $\bar A_\3$ and $\bar
A_\2$ defined by
\be
\hat A_\3 = \bar A_\3 + \bar A_2\wedge (dz+\cA_\1)\,,\label{a3a2}
\ee
we can read off the As\"atze for $\bar A_\3$ and $\bar A_\2$.
To do this, the following {\it lemmata} are useful:
\bea
\ep_\3&=& \ft1{\sqrt2}\, g\, \ep_{ijk}\, \mu_\k \, D\mu^i\wedge D\mu^j
\wedge (dz+ \cA_\1)\,,\nn\\
F_\2^i\wedge h_i &=& -F_\2^{ij}\wedge (\mu^i\, D\mu^j) + \ft1{\sqrt2}\,
g\, \ep_{ijk}\, \mu^k\, F_\2^{ij}\wedge(dz+\cA_\1)\,.\label{intred}\\
\ft12\ep_{ijk}\, F_\2^i\wedge h_j\wedge h_k &=& \ft12 F_\2^{ij}\wedge
D\mu^i\wedge D\mu^j - \ft1{\sqrt2}\, g\, \ep_{ijk}\,
F_\2^{ij} \wedge D\mu^k\wedge (dz+\cA_\1)\,.\nn
\eea

    From these results, it then follows that the Ans\"atze for the
ten-dimensional 3-form and 2-form potentials defined by (\ref{a3a2}) are
\bea
\bar A_\3 &=& s\, A_\3 +\ft1{\sqrt2}\, g^{-2}\,s\, F_\2^{ij}\wedge
(\mu^i\, D\mu^j) - \ft1{\sqrt2}\, g^{-1}\, s\, \omega_\3\,,\nn\\
\bar A_\2 &=&  \ft1{4}\, g^{-3}\, s\, \mu^k\, \ep_{ijk}\, \Big(
(2 + c^2\, X^{-4}\, \Delta^{-1})\, D\mu^i\wedge D\mu^j - 2g\,
F_\2^{ij}\Big)\,.\label{a3a2exp}
\eea
Finally, note that the expression (\ref{kkvector}) for the
Kaluza-Klein vector $\cA_\1$ can be shown, after some algebra, to
imply the following expression for the corresponding field strength
$\cF_\2\equiv d\cA_\1$:
\be
\cF_\2 = \ft1{2\sqrt2}\, g^{-1}\, \ep_{ijk}\, \mu^k\, D\mu^i \wedge
D\mu^j -\ft1{2\sqrt2}\, \ep_{ijk}\, \mu^k\, F_\2^{ij}\,.
\ee

    In fact this expression for $\cF_\2$ allows a rewriting of $\bar
A_\2$, given in (\ref{a3a2exp}), in a slightly more elegant way:
\be
\bar A_\2 = \sqrt2\, g^{-2}\, s\, \cF_\2 +\ft1{4}\, g^{-3}\, s\, c^2\,
X^{-4}\, \Delta^{-1}\, \ep_{ijk}\, \mu^k\, D\mu^i\wedge D\mu^j\,.
\ee

\section{$D=6$ $SU(2)$-gauged $N=2$ supergravity from type IIB}

   The embedding of six-dimensional $N=2$ $SU(2)$-gauged supergravity
in the massive type IIA theory was obtained in \cite{d6gauge}.  Here,
we apply a similar $U(1)$ Hopf reduction to this local $S^4$
reduction, thereby obtaining a reduction of $D=9$ supergravity to the
$D=6$ gauged theory.  Then, by applying the standard T-duality rules,
we can lift the nine-dimensional theory back to $D=10$, expressed now
as a consistent reduction of type IIB supergravity.

   First, consider the metric and dilaton $\phi_1^{IIA}$.  From
\cite{d6gauge} we have
\bea
d\hat s_{10}^2(IIA) &=& s^{1/12}\, X^{1/8}\, \Big[ \Delta^{3/8}\,
ds_6^2 + 2 g^{-2}\, X^2\, \Delta^{3/8}\, d\xi^2 + \ft12 g^{-2}\,
X^{-1}\, \Delta^{-5/8}\, c^2\, \sum_i h_i^2\Big]\,,\nn\\
e^{\phi_1^{IIA}} &=& s^{-5/6}\, X^{-5/4}\, \Delta^{1/4}\,,\label{2aans}
\eea
where in this case 
\be
\Delta= X\, c^2 + X^{-3}\, s^2\,.
\ee
The reduction to $D=9$ is as follows:
\be
d\hat s_{10}^2(IIA) = e^{-2\a\, \phi_2^{IIA}}\, ds_9^2 + e^{14\a\,
\phi_2^{IIA}}\, (dz + \cA_\1^{IIA})^2\,,\label{2ared}
\ee
where $\a=1/(4\sqrt7)$.  
From (\ref{hexp0}) the internal metric can be rewritten using
\be
\sum_i h_i^2 = \sum_i (D\mu^i)^2 + 2g^2\, (dz + \cA_\1^{IIA})^2\,.
\label{hexp}
\ee

    Substituting (\ref{hexp}) into (\ref{2aans}), and comparing with
(\ref{2ared}), we therefore obtain the following expressions for the
nine-dimensional metric and dilatons:
\bea
ds_9^2 &=& s^{2/21}\, c^{2/7}\,  \Big[ \Delta^{2/7}\,
ds_6^2 + 2 g^{-2}\, X^2\, \Delta^{2/7}\, d\xi^2 + \ft12 g^{-2}\,
X^{-1}\, \Delta^{-5/7}\, c^2\, D\mu^i\, D\mu^i\Big]\,,\nn\\
e^{\phi_1^{IIA}} &=& s^{-5/6}\, X^{-5/4}\,
\Delta^{1/4}\,,\label{2ain9}\\
e^{2\a\, \phi_2^{IIA}} &=& s^{1/84}\, c^{2/7}\, X^{-1/8}\,
\Delta^{-5/56}\,.
\eea

     Under the IIA/IIB T-duality transformation in $D=9$, the IIB
dilatons are related to the IIA dilatons by the orthogonal
transformation
\be
\phi_1^{IIB} = \ft34\, \phi_1^{IIA} -\ft{\sqrt7}{4}\,
\phi_2^{IIA}\,,\qquad
\phi_2^{IIB} = -\ft{\sqrt7}{4}\, \phi_1^{IIA} - \ft34\,
\phi_2^{IIA}\,.\label{2a2bdilatons}
\ee
The nine-dimensional metric and dilatons can now be lifted back to ten
dimensions in the IIB variables, using the analogue of (\ref{2ared}),
namely
\be
d\hat s_{10}^2(IIB) = e^{-2\a\, \phi_2^{IIB}}\, ds_9^2 + e^{14\a\,
\phi_2^{IIB}}\, (dz + \cA_\1^{IIB})^2\,,\label{2bred}
\ee
Doing this, we obtain 
\bea
d\hat s_{10}^2(IIB) &=& c^{1/2}\, X^{-1/4}\, \Big[\Delta^{1/4}\,
ds_6^2 + 2 g^{-2}\, X^2\, \Delta^{1/4}\, d\xi^2 + \ft12 g^{-2}\,
X^{-1}\, \Delta^{-3/4}\, c^2\, D\mu^i\, D\mu^i\Big]\nn\\
&&+ s^{2/3}\, c^{-3/2}\, X^{7/4}\, \Delta^{1/4}\, (dz+ \cA_\1^{IIB})^2
\,,\nn\\
e^{\phi_1^{IIB}} &=& s^{-2/3}\, c^{-1}\, X^{-1/2}\,
\Delta^{1/2}\,.\label{2bmet1} 
\eea
(As we shall see below, after making the $S^1$ reduction of the field
strengths, the Kaluza-Klein vector $\cA_\1^{IIB}$ in the type IIB
picture, which comes from the winding vector of the original type IIA
description, is actually zero.)  The vacuum solution, corresponding to
taking $X=1$, $A_\1^{ij}=0$ and $ds_6^2$ to be AdS$_6$, can be
obtained as the near-horizon limit of a semi-localised D5/D7/NS5-brane
system \cite{clpv}.

     It is interesting to note that if we express the two
ten-dimensional metrics in their respective string frames, related to
the Einstein frames by
\be
ds_{10}^2(str) = e^{\fft12\phi_1}\, d\hat s_{10}^2\,,\label{stringmet}
\ee
then we get the following:
\bea
ds_{10}^2(IIA,str) &=& s^{-1/3} \, X^{-1/2} \, \Big[ \Delta^{1/2}\,
ds_6^2 + 2 g^{-2}\, X^2\, \Delta^{1/2}\, d\xi^2 \nn\\
&& + \ft12 g^{-2}\,
X^{-1}\, \Delta^{-1/2}\, c^2\, D\mu^i\, D\mu^i\Big] 
+s^{-1/3}\,  X^{-3/2}\, c^2\, (dz+
\cA_\1^{IIA})^2\,,\nn\\
ds_{10}^2(IIB,str) &=& s^{-1/3} \, X^{-1/2} \, \Big[ \Delta^{1/2}\,
ds_6^2 + 2 g^{-2}\, X^2\, \Delta^{1/2}\, d\xi^2 \nn\\
&&+ \ft12 g^{-2}\,
X^{-1}\, \Delta^{-1/2}\, c^2\, D\mu^i\, D\mu^i\Big] 
+s^{1/3}\,  X^{3/2}\, c^{-2}\, (dz+
\cA_\1^{IIB})^2\,.\label{2bmet2}
\eea
Thus we see that the effect of the T-duality is, as one might expect,
simply to invert the prefactor in the $U(1)$ direction.

    From \cite{d6gauge}, the Ans\"atze for the reduction to $D=6$ of the
various field strengths of the massive type IIA theory are
\bea
\hat F_\4 &=& -\ft{\sqrt2}{6}\, g^{-3}\, s^{1/3}\, c^3\, \Delta^{-2}\,
U\, d\xi\wedge\ep_\3 -\sqrt2 g^{-3}\, s^{4/3}\, c^4\, \Delta^{-2}\,
X^{-3}\, dX\wedge \ep_\3 \nn\\
&&-\sqrt2 g^{-1}\,
s^{1/3}\, c\, X^4\, {*F_\3}\wedge d\xi
-\ft1{\sqrt2} s^{4/3}\, X^{-2}\, {*F_\2} \nn\\
&& +\ft1{\sqrt2} g^{-2}\,
s^{1/3}\, c\, F_\2^i \wedge h^i\wedge d\xi -\ft1{4\sqrt2} g^{-2}\,
s^{4/3}\, c^2\, \Delta^{-1}\, X^{-3}\,  F_\2^i \wedge
h^j\wedge h^k\, \ep_{ijk}\,,\label{fans}\\
\hat F_\3 &=& s^{2/3}\, F_\3 + g^{-1}\, s^{-1/3}\, c\, F_\2\wedge d\xi
\,,\nn\\
\hat F_\2 &=& \ft1{\sqrt2}\, s^{2/3}\, F_\2\,,\qquad
e^{\hat\phi} = s^{-5/6}\, \Delta^{1/4}\, X^{-5/4}\,,\nn
\eea
In the Hopf reduction from $D=10$ to $D=9$, we follow the standard
Kaluza-Klein rules, with the field strengths reduced as follows:
\bea
\hat F_\4 &=& \bar F_\4 + \bar F_{\3 1}\wedge (dz + \cA_\1^{IIA})\,,\nn\\
\hat F_\3 &=& \bar F_\3 + \bar F_{\2 1}\wedge (dz + \cA_\1^{IIA})\,,\\
\hat F_\2 &=& \bar F_\2 + \bar F_{\1 1}\wedge (dz + \cA_\1^{IIA})\,,\nn
\eea

    Using the {\it lemmata} given in (\ref{intred}), it is now
straightforward to read off the expressions for the nine-dimensional
fields:
\bea
\bar F_\4 &=& -\sqrt2\, g^{-1}\, s^{1/3}\, c\, X^4\, {*F_\3} \wedge
d\xi -\ft1{\sqrt2}\, s^{4/3}\, X^{-2}\, {*F_\2} \nn\\
&& -\ft1{\sqrt2}\, g^{-2}\, s^{1/3}\, c\, F_\2^{ij} \wedge (\mu^i\,
D\mu^j)\wedge d\xi - \ft1{4\sqrt2}\, g^{-2}\, s^{4/3}\, c^2\, X^{-3}\,
\Delta^{-1}\, F_\2^{ij}\wedge D\mu^i\wedge D\mu^j\,,\nn\\
\bar F_{\3 1} &=& -\ft16 g^{-2}\, s^{1/3}\, c^3\, \Delta^{-2}\, U\,
\ep_{ijk}\, \mu^k\, D\mu^i\wedge D\mu^j\wedge d\xi \nn\\
&&- g^{-2}\,
s^{4/3}\, c^4\, X^{-3}\, \Delta^{-2}\, \ep_{ijk}\, \mu^k\, dX\wedge
D\mu^i\wedge D\mu^j \nn\\
&&-\ft12 g^{-1}\, s^{1/3}\, c\, \mu^k\, \ep_{ijk}\, F_\2^{ij}\wedge
d\xi + \ft14 g^{-1}\, s^{4/3}\, c^2\, X^{-3}\, \Delta^{-1}\,
\ep_{ijk}\, F_\2^{ij}\wedge D\mu^k\,,\nn\\
\bar F_\3 &=& s^{2/3}\, F_\3 + g^{-1}\, s^{-1/3}\, c\, F_\2\wedge d\xi
\,,\nn\\
\bar F_{\2 1} &=& 0\,,\label{d9fields}\\
\bar F_\2 &=& \ft1{\sqrt2}\, s^{2/3}\, F_\2\,,\nn\\
\bar F_{\1 1} &=& 0\,.\nn
\eea 

   From these, we may note the following.  Firstly, the fact that the
NS-NS field $\bar F_{\2 1}$ is zero means that after the T-duality
transformation, which maps this into the Kaluza-Klein 2-form field
strength $\cF_\2^{IIB}\equiv d\cA_\1^{IIB}$ of the type IIB reduction
to $D=9$, we shall have $\cA_\1^{IIB}=0$.  This means that in the
expressions in (\ref{2bmet1}) and (\ref{2bmet2}) for the type IIB
metric, the contribution in the $z$ direction involves just a pure
untwisted $dz^2$.

    After lifting the various nine-dimensional fields given in
(\ref{d9fields}) to $D=10$ in the type IIB variables (see, for example
\cite{bergs,clps}), we therefore find that the Ans\"atze for the self-dual
5-form, and the R-R and NS-NS three forms, are
\bea
\wtd F_\5 &=& {* \bar F_\4} + \bar F_\4 \wedge dz\,,\nn\\
\wtd F_\3^{RR} &=& - \bar F_{\3 1} + \bar F_\2 \wedge dz\,,\\
\wtd F_\3^{NS} &=&\bar F_\3 + \cF_\2^{IIA}\wedge dz\,.\nn
\eea
Since $\bar F_{\1 1}$ is zero, there is no axionic field excitation.
However, since the T-duality that relates the massive IIA theory to
the type IIB theory involves a generalised Scherk-Schwarz reduction
\cite{green}, the ten-dimensional axion $\hat \chi$ of the type IIB theory
is given by
\be
\hat \chi = m\, z
\ee
in the reduction to gauged six-dimensional supergravity.

\section{$D=5$ $SU(2)\times U(1)$ gauged $N=4$ supergravity from type
IIA and $D=11$}

    Five-dimensional $N=4$ $SU(2)\times U(1)$-gauged supergravity was
obtained as a consistent $S^5$ reduction of type IIB supergravity in
\cite{d5gauge}.  The 5-sphere can be viewed as a foliation of
$S^3\times S^1$ surfaces.  In this section, we perform an $S^1$
reduction on the $U(1)$ Hopf fibres over the $S^3$, thereby obtaining
a reduction Ansatz for a nine-dimensional embedding of the
five-dimensional theory.  After a T-duality transformation, we can
then express this as a consistent embedding of the five-dimensional
gauged supergravity in the type IIA theory, and eleven-dimensional
supergravity.

    The Kaluza-Klein $S^5$ reduction Ansatz from the type IIB theory
is given by \cite{d5gauge}:
\bea
d\hat s_{10}^2 &=& \Delta^{1/2}\, ds_5^2 + 2 g^{-2}\, \Delta^{1/2}\, X\,
d\xi^2 + \Delta^{-1/2}\, X^2\, s^2\, (d\tau + B_\1)^2 \nn\\
&&\qquad + \ft12 g^{-2}\,
\Delta^{-1/2}\, X^{-1}\, c^2 \, \sum_i h_i^2\,,\nn\\
\hat{G}_\5 &=& \sqrt2\, g\, U \,\varepsilon_5 - \frac{3\sqrt2\,
sc}{g} X^{-1}\,
{*dX}\wedge d\xi +
\frac{c^2}{4\sqrt{2}\, g^2} X^{-2}\, {*F^i_\2}\wedge h^j\wedge h^k
\, \varepsilon_{ijk} \nn\\
& &-\frac{sc}{\sqrt{2}\,g^2} X^{-2}\, {*F^i_\2}\wedge h^i\wedge d\xi
+ \frac{2sc}{g^2} X^4\, {*G_\2}\wedge d\xi\wedge (d\tau + g B_\1),\nn\\
\hat{A}_\2 &\equiv& \hat A_\2^{RR} + \im\, \hat A_\2^{NS} =
 -s\, g^{-1}\, e^{\im\,g\, \tau/\sqrt2}\, A_\2\,,\nn\\
\hat{\phi} &=& 0,\ \ \ \hat{\chi} = 0,
\label{10metans}
\eea
where the self-dual 5-form is given by $\hat H_\5 = \hat G_\5 + {\hat
* \hat G_\5}$, $U\equiv X^2\, c^2 +
X^{-1}\, s^2 + X^{-1}$, and $\ep_5$ is the volume form in the
five-dimensional spacetime metric $ds_5^2$, and
\be
\Delta = X\, c^2 + X^{-2}\, s^2\,.
\ee
Note that we have defined the complex 2-form potential $\hat A_\2\equiv
\hat A_\2^1 + \im\, \hat A_\2^2$ in the type IIB theory.
The ten-dimensional dilaton and the axion are constants,
which without loss of generality we have set to zero.
The conventions that we are using here are related to those in
\cite{d5gauge} by making the following replacements on the quantities
in \cite{d5gauge}: $g\longrightarrow g/\sqrt2$, $\tau\longrightarrow
-g\, \tau/\sqrt2$.  Note that the scalar $X$ and the gauge fields
$A_\1^{ij}$ parameterise deformations of a 5-sphere.  It is foliated
by $S^3\times S^1$, where $\tau$ is the coordinate on the $S^1$ factor.

    Following an analogous strategy to that of the previous section,
we substitute (\ref{hexp0}) and (\ref{intred}) into the Ansatz, and
perform a T-duality transformation on the $z$ coordinate.   We find
that in the string frame, the resulting type IIA metric Ansatz is
given by
\bea
ds_{10}^2(IIA,str) 
&=& \Delta^{1/2}\, ds_5^2 + 2 g^{-2}\, \Delta^{1/2}\, X\,
d\xi^2 + \Delta^{-1/2}\, X^2\, s^2\, (d\tau + B_\1)^2\nn\\
&& + \ft12 g^{-2}\,
\Delta^{-1/2}\, X^{-1}\, c^2 \, \sum_i (D\mu^i)^2
 + \Delta^{1/2}\, X\, c^{-2}\, dz_2^2\,,
\eea
and the dilaton of the type IIA theory is given by
\be
e^{\phi_1^{IIA}} = \Delta^{1/4}\, X^{1/2}\, c^{-1}\,.
\ee
(We are naming the reduction coordinate $z_2$ here, in anticipation of
performing a further oxidation to $D=11$ presently.)
    
   The field strengths of the type IIA theory turn out to be as
follows:
\bea
F_\4^{IIA} &=& \wtd F_\4 - F_\3^{RR}\wedge dz_2\,,\nn\\
F_\3^{IIA} &=& F_\3^{NS}  + \cF_\2^{IIB}\wedge dz_2 \,,\\
F_\2^{IIA} &=& 0\,,\nn
\eea
where
\bea
&&\wtd F_\4 = \Big[ \ft12 s\, c^3\, g^{-2}\, U\, \Delta^{-2}\,
\mu^k\, d\xi\wedge D\mu^i\wedge D\mu^j \wedge (d\tau+B_\1) \nn\\
&&\qquad 
-\ft34 s^2\, c^4\, g^{-2}\, \Delta^{-2}\, X^{-2}\, \mu^k\, dX\wedge D\mu^i
\wedge D\mu^j \wedge (d\tau+B_\1) \nn\\
&&\qquad +\ft1{4\sqrt2}\, g^{-1}\, s^2\, c^2 \, \Delta^{-1}\, X^{-2}\,
F_\2^{ij}\wedge D\mu^k\wedge (d\tau+B_\1) \nn\\
&&\qquad -\ft1{2\sqrt2}\, g^{-1}\, s\, c\,
\mu^k\, d\xi\wedge F_\2^{ij}\wedge (d\tau+B_\1)
- \ft14 g^{-1}\, c^2\, X^{-2}\, {*F_\2^{ij}} \wedge D\mu^k \nn\\
&&
\qquad + \ft12 g^{-1}\, s\, c\, \mu^k\, X^{-2}\, {*F_\2^{ij}} \wedge d\xi
-\ft14 g^{-2}\, c^4\, \mu^k\, \Delta^{-1}\,  X\, G_\2\wedge D\mu^i
\wedge D\mu^j\Big]\, \ep_{ijk}\,,\nn\\ 
&&F_\3^{RR} + \im\, F_\3^{NS} = d\hat A_\2\,,\\
&&\cF_\2^{IIB} =\ft1{2\sqrt2}\, g^{-1}\, \ep_{ijk}\, \mu^k\, D\mu^i \wedge
D\mu^j -\ft1{2\sqrt2}\, \ep_{ijk}\, \mu^k\, F_\2^{ij}\,.\nn
\eea

    The embedding of $D=5$ $SU(2)\times U(1)$ gauged $N=4$
supergravity in type IIA supergravity that we have just derived can be
lifted further, to $D=11$ supergravity.  For the eleven-dimensional
metric, we find
\bea
d\hat s_{11}^2 &=& X^{-1/3}\, c^{2/3}\, \Big[
\Delta^{1/3}\, ds_5^2 + 2 g^{-2}\, \Delta^{1/3}\, X\,
d\xi^2 + \Delta^{-2/3}\, X^2\, s^2\, (d\tau + B_\1)^2\nn\\
&& + \ft12 g^{-2}\,
\Delta^{-2/3}\, X^{-1}\, c^2 \, \sum_i (D\mu^i)^2\Big] 
 + \Delta^{1/3}\, X^{2/3}\, c^{-4/3}\, (dz_1^2+ dz_2^2)\,,
\eea
where $z_1$ is the coordinate on the additional $S^1$.  The 4-form
field strength in $D=11$ is given by
\be
\hat F_\4 = \wtd F_\4 + {\cal I}\!m [\wtd F_\3\wedge (dz_1 - \im\, dz_2)]
 -\cF_\2^{IIB} \wedge dz_1 \wedge dz_2\,.
\ee

   The vacuum solution, corresponding to setting $X=1$, $A_\1^{ij}=0$,
$B_\1=0$ and taking $ds_5^2$ to be AdS$_5$, can be viewed as the
near-horizon limit of a semi-localised M5/M5-brane system \cite{preoz2,oz2}.
Note that the Hopf T-duality has the effect of untwisting the $S^3$
into $S^2\times S^1$.  The effect of this procedure on AdS$_3\times
S^3$ was extensively studied in \cite{ads3s3}.

\section{$SO(4)$-gauged $N=4$ supergravity in $D=4$ from type IIB}

     The $SO(4)$-gauged $N=4$ supergravity in $D=4$ was explicitly
obtained as an $S^7$ reduction from $D=11$ supergravity
\cite{d4gauge}.  In this reduction, the $S^7$ has a natural
description in terms of a foliation of $S^3\times S^3$ surfaces.  The
two copies of $SU(2)$ in the $SO(4)\sim SU(2)\times SU(2)$ gauge group
come from left-invariant actions on the two copies of $S^3$.  Since
there are two Hopf circles, one from each $S^3$, we can perform two
steps of Kaluza-Klein $S^1$ reduction.  The first gives an embedding
of the $N=4$ gauged theory in type IIA supergravity, and the second,
combined with a T-duality transformation, gives the embedding of the
$N=4$ gauged theory in type IIB supergravity.

    Since the expression in \cite{d4gauge} for the Kaluza-Klein $S^7$
reduction of the $D=11$ 4-form is very complicated, we shall not
present explicit formulae here for the field strengths in the type IIA
and type IIB pictures.  It is completely straightforward to obtain
them, by following the same steps as we did in previous sections.
Thus we shall just present the Kaluza-Klein Ansatz for the metric
reductions here.

     The Kaluza-Klein Ansatz for the reduction of the
eleven-dimensional metric is \cite{d4gauge}
\be
d\hat s_{11}^2 = \Delta^{\fft23}\, ds_4^2 + 2 g^{-2}\,
\Delta^{\fft23}\, d\xi^2
+ \ft12
g^{-2}\, \Delta^{\fft23}\, \Big[ c^2\, \Omega^{-1}\, \sum_i (h^i)^2
+ s^2\, \wtd\Omega^{-1}\, \sum_i (\td h^i)^2\Big]\,,\label{metans}
\ee
where
\bea
&&\tX \equiv  X^{-1}\, q\,,\qquad q^2 \equiv 1 + \chi^2\,
X^4\,,\nn\\
&&\Omega\equiv c^2\, X^2 + s^2\,,\qquad \wtd\Omega\equiv s^2\, \wtd
X^2 + c^2\,,\nn\\
&&\Delta \equiv  \Big[(c^2\, X^2 + s^2)(s^2\, \tX^2 + c^2)
\Big]^{\fft12} \,,\label{defs1}\\
&&c\equiv \cos\xi\,,\qquad s\equiv \sin\xi\,,\nn\\
&& h^i \equiv \sigma_i - g\, A_\1^i\,,\qquad \td h^i \equiv \td\sigma_i
-g\, \wtd A_\1^i\,.\nn
\eea 
Here $X=e^{\fft12\phi}$, and $(\phi,\chi)$ are the dilatonic and
axionic scalars of the four-dimensional gauged theory.

   As a first step, we make a Hopf reduction on the untilded $S^3$,
using the expression (\ref{hexp0}).  Comparing with the standard $S^1$
reduction in (\ref{d11d10red}), this gives the type IIA ten-dimensional
metric and dilaton:
\bea
ds_{10}^2(IIA)&=& \Delta^{3/4}\, c^{1/4}\, \Omega^{-1/8}\, 
\Big[ ds_4^2 + 2g^{-2}\, d\xi^2 + 
\ft12 g^{-2}\, c^2\, \Omega^{-1}\, \sum_i (D\mu^i)^2\nn\\
&&\qquad\qquad\qquad
+ \ft12 g^{-2}\, s^2\, \wtd\Omega^{-1}\, \sum_i (\td h^i)^2
\Big]\,,\nn\\
e^{\phi_1^{IIA}} &=& \Delta^{1/2}\,  c^{3/2}\, \Omega^{-3/4} \,.
\eea

   The next step is to perform a Hopf reduction on the second $S^3$
factor.  Denoting all relevant quantities with tildes, we use the same
result (\ref{hexp0}), and reduce the metric according to the standard
$S^1$ reduction (\ref{2ared}).  This gives the nine-dimensional
metric, and second dilaton:
\bea
ds_9^2(IIA) &=& \Delta^{4/7}\, (s\, c)^{2/7} \, \Big[
 ds_4^2 + 2g^{-2}\, d\xi^2 + 
\ft12 g^{-2}\, c^2\, \Omega^{-1} \, \sum_i (D\mu^i)^2\nn\\
&&\qquad\qquad\qquad
+ \ft12 g^{-2}\, s^2\, \wtd\Omega^{-1}\, \sum_i (\wtd D\td\mu^i)^2
\Big]\,,\nn\\
e^{2\a\, \phi_2^{IIA}} &=& \Delta^{3/28}\,
\Big(\fft{c^2}{\Omega}\Big)^{1/56}
\, \Big(\fft{s^2}{\wtd\Omega}\Big)^{1/7}\,.
\eea

   After transforming to type IIB variables, including the dilaton
transformation (\ref{2a2bdilatons}), the metric can be oxidised back
to $D=10$, as an embedding now in the type IIB theory.  This metric,
and the corresponding type IIA metric before the Hopf T-duality
transformation, are most usefully expressed in the string frame.  
The expressions are as follows:
\bea
ds_{10}^2(IIA,str) &=& 
\Delta\, c\, \Omega^{-1/2}\, 
\Big[ds_4^2 + 2g^{-2}\, d\xi^2 + 
\ft12 g^{-2}\, c^2\, \Omega^{-1} \, \sum_i (D\mu^i)^2\nn\\
&&\qquad
+ \ft12 g^{-2}\, s^2\, \wtd\Omega^{-1}\, \sum_i (\wtd D\td\mu^i)^2
\Big] + s^{2}\, c^{1}\, \wtd\Omega^{-1/2}\, (dz_2+\cA_\1)^2\,,\nn\\
ds_{10}^2(IIB,str) &=& 
\Delta\, c\, \Omega^{-1/2}\, 
\Big[ds_4^2 + 2g^{-2}\, d\xi^2 + 
\ft12 g^{-2}\, c^2\, \Omega^{-1} \, \sum_i (D\mu^i)^2\nn\\
&&\qquad
+ \ft12 g^{-2}\, s^2\, \wtd\Omega^{-1}\, \sum_i (\wtd D\td\mu^i)^2
\Big] + s^{-2}\, c^{-1}\, \wtd\Omega^{1/2}\, dz_2^2\,.
\eea

     Note that there is no ``twist'' involving the $z_2$ coordinate in
the type IIB ten-dimensional metric.  This is a reflection of the fact
that there is no winding vector in $D=9$ in the type IIA reduction.
Such a vector would have come from the reduction of the 3-form field
strength in the $D=10$ type IIA theory.  This, in turn comes from the
reduction of the 4-form of $D=11$.  But the 3-form in $D=10$ comes
from the Hopf reduction of the first $S^3$ factor in the $S^3\times
S^3$ foliation of $S^7$.  Consequently, it has no terms involving the
directions in the second $S^3$ factor, and so no winding vector
emerges in $D=9$.

    For a similar reason, the axion $\bar\chi$ of the type IIB theory
is zero in the reduction Ansatz.  It would correspond, in the type IIA
picture, to the axion that would come from the reduction of the
Kaluza-Klein vector in $D=10$.  But this lives in the directions of
the first $S^3$ (see (\ref{kkvector})), and so it does not give rise
to any axion when the further reduction to $D=9$ on the Hopf fibres of
the second $S^3$ is performed.

    The vacuum solution, corresponding to taking $X=1$, $\chi=0$,
$A_\1^{ij}=0$, $\wtd A_\1^{ij}=0$ and $ds_4^2$ to be AdS$_4$, can be
viewed as the near-horizon limit of a semi-localised D2/D6-brane
system in type IIA supergravity or a D3/D5/NS5-system in type IIB
supergravity \cite{clpv}.

\section{Hopf reduction on non-singular fibres}

     All the examples that we have considered so far in this paper
involve performing Hopf reductions on circles whose radius goes to
zero for some value of the azimuthal coordinate $\xi$ on the internal
spherical manifold.  For example, when we reinterpreted the $S^4$
reduction of $D=11$ supergravity in section 2 as a reduction of type
IIA supergravity, the circle parameterised by $z$ in
(\ref{d11metred2}) reduced to zero radius at $\xi=\ft12\pi$.  In
certain cases a more general type of Hopf reduction can be performed,
in which the radius of the $U(1)$ fibres remains non-zero for all
values of $\xi$.  Specifically, this can be done for the $S^5$ and
$S^7$ reductions.  The reason for this is that in each of these
examples, there are in fact two $U(1)$ Killing vectors in the
higher-dimensional metric, corresponding to the fact that the foliating
surfaces at constant $\xi$ are the product of two odd-dimensional
spheres ($S^3\times S^1$ and $S^3\times S^3$ respectively).  In each
case when the radius of one of the spheres goes to zero (at $\xi=0$ or
$\xi=\ft12\pi$), the other has non-zero radius.  Thus by taking a
linear combination of the two Killing directions for the $S^1$
Kaluza-Klein reduction, a non-singular embedding can be achieved.

   To see how this works, consider the relevant two-dimensional 
factor in the higher-dimensional metric, namely the part involving the
two $U(1)$ directions.  We shall write this as
\be
d\bar s^2 = \a^2 \, (d\tau_1 +\cA^1_\1)^2 + \beta^2\, (d\tau_2
+\cA_\1^2)^2\,.
\ee
Now, we make the coordinate redefinitions
\be
\tau_1 = a\, x+ b\, y\,,\qquad \tau = - b\, x + a\, y\,,\label{lincom}
\ee
where $a$ and $b$ are constants.
It is straightforward to establish the following lemma:
\bea
&&d\bar s^2 =
(\a^2\, b^2 + \beta^2\, a^2)\, \Big[dy + \fft{a\, b\, 
(\a^2-\beta^2)\, dx + \a^2\, b\, \cA_\1^1 + \beta^2\, a\,
\cA_\1^2}{\a^2\, b^2 + \beta^2\, a^2}\Big]^2 \nn\\
&&\qquad\qquad \qquad+ \fft{\a^2\,
\beta^2}{\a^2\, b^2 + \beta^2\, a^2}\, \Big[(a^2+b^2)\, 
dx + a\, \cA_\1^1 - b\, \cA_\1^2\Big]^2\,.\label{lincomlem}
\eea
If we now perform an $S^1$ reduction on the $y$ coordinate we see
that the $U(1)$ fibres will always have non-zero length, provided that
the functions $\a$ and $\beta$ do not vanish simultaneously.  Since,
in both our examples one of them is proportional to $\sin\xi$, while
the other is proportional to $\cos\xi$, this condition for
non-singularity of the $y$ fibres is satisfied.

\subsection{Non-singular Hopf reduction for the $S^5$ embedding}

    Using (\ref{hexp0}), the metric Ansatz for the $S^5$ reduction
given in (\ref{10metans}) can be rewritten as
\bea
d\hat s_{10}^2 &=& \Delta^{1/2}\, ds_5^2 + 2 g^{-2}\, \Delta^{1/2}\, X\,
d\xi^2 + \Delta^{-1/2}\, X^2\, s^2\, (d\tau + B_\1)^2 \nn\\
&&+ \ft12 g^{-2}\,
\Delta^{-1/2}\, X^{-1}\, c^2 \, \sum_i (D\mu^i)^2 +
\Delta^{-1/2}\, X^{-1}\, c^2\, (dz+\cA_\1)^2\,.\label{d5case1}
\eea
As far as this metric reduction Ansatz is concerned, we see that
there are two $U(1)$ Killing directions, namely $z$ and $\tau$.
Accordingly, we can choose a more general reduction scheme, in which
we take a linear combination of these two coordinates for our circle
reduction.  

   Comparing (\ref{lincomlem}) with (\ref{d5case1}), we see that the
functions $\a$ and $\beta$ are given by
\be
\a^2 = \Delta^{-1/2}\, X^{-1}\, c^2 ,,\qquad \beta^2 =\Delta^{-1/2}\,
X^2\, s^2\,.
\ee
Reducing on the $y$ coordinate, following the standard procedure, we
arrive at the nine-dimensional metric
\bea
ds_9^2 &=& \wtd\Delta^{1/7}\, \Delta^{3/7}\, \Big[ ds_5^2 + 2 g^{-2}\,
X\, d\xi^2 +\ft12 g^{-2}\, \Delta^{-1}\, X^{-1}\, c^2\, \sum_i
(D\mu^i)^2 \nn\\
&&\qquad+ \wtd\Delta^{-1}\, \Delta^{-1}\, X\, s^2\, c^2\,
[(a^2+b^2)\, 
dx +a\, \cA_\1 - b\, B_\1]^2\Big]\,,
\eea
where we have defined 
\be
\wtd\Delta = b^2\, X^{-1}\, c^2 + a^a\, X^2\, s^2\,.
\ee
Note that the dilaton $\phi_2^{IIB}$ resulting from the Kaluza-Klein
reduction from $D=10$ to $D=9$ is given by
\be
e^{2\a\, \phi_2^{IIB}} = \Delta^{-1/14}\, \wtd\Delta^{1/7}\,,
\ee
where as usual $\a=1/(4\sqrt7)$.  

     Since the type IIB dilaton $\phi_1^{IIB}$ in the original $S^5$
reduction is zero, it follows from (\ref{2a2bdilatons}) that after
performing a T-duality transformation in $D=9$, and lifting back to
the type IIA theory, the metric becomes
\bea
d\hat s_{10}^2(IIA) &=& 
\wtd\Delta^{1/4}\, \Delta^{3/8}\, \Big[ ds_5^2 + 2 g^{-2}\,
X\, d\xi^2 +\ft12 g^{-2}\, \Delta^{-1}\, X^{-1}\, c^2\, \sum_i
(D\mu^i)^2 \nn\\
&&\qquad + \wtd\Delta^{-1}\, \Delta^{-1}\, X\, s^2\, c^2\, [(a^2+b^2)\,
dx +a\, \cA_\1 -b\,  B_\1]^2 +  \wtd\Delta^{-1}\, dy^2\Big] \,.
\label{d5met3}
\eea
(Note that as usual there is no longer any ``twist'' in the $y$
direction of the $S^1$, after the T-duality transformation.) 
This embedding can be further lifted to $D=11$, with the metric given
by
\bea
d\hat s_{11}^2 &=& 
(\wtd\Delta\, \Delta)^{1/3}\, \Big[ ds_5^2 + 2 g^{-2}\,
X\, d\xi^2 +\ft12 g^{-2}\, \Delta^{-1}\, X^{-1}\, c^2\, \sum_i
(D\mu^i)^2 \nn\\
&&\qquad + \wtd\Delta^{-1}\, \Delta^{-1}\, X\, s^2\, c^2\, [(a^2+b^2)\,
dx +a\, \cA_\1 -b\,  B_\1]^2 +  \wtd\Delta^{-1}\, (dz_1^2+dy^2)\Big] \,,
\label{d5met4}
\eea
where $z_1$ is the eleventh coordinate.

    It is instructive to examine the metric (\ref{d5met4}) in the
``vacuum'' case where the lower-dimensional scalar and gauge fields
are set to zero, in which case we can take $ds_5^2$ to be AdS$_5$.
Bearing in mind that $\cA_\1$ is still non-zero, and given by the
first term in (\ref{kkvector}), we see that the eleven-dimensional metric
becomes
\be
d\hat s_{11}^2 = \wtd\Delta^{1/3}\, ds_5^2 + 2 g^{-2}\,
\wtd\Delta^{1/3}\, d\bar s^2 + \wtd\Delta^{-2/3}\, (dz_1^2 + dy^2)\,,
\ee
where
\bea
d\bar s^2 &=& d\xi^2 + \ft14 c^2\, (d\theta^2 + \sin^2\theta\, d\psi^2)
+ \ft14 s^2\, c^2\, (d\sigma + \cos\theta\, d\psi)^2\,,\nn\\
\wtd \Delta &=& b^2\, c^2 + a^2\, s^2\,,\label{ads5nonsing}\\
\sigma &=&  \fft{\sqrt 2\, g\, (a^2+b^2)}{a}\, x\,.\nn
\eea
In order for the level surfaces at constant $\xi$ to be globally
non-singular, the angular coordinate $\sigma$ should be chosen to have period
$4\pi/p$, where $p$ is an integer.  The level surfaces are then cyclic
lens spaces $S^3/Z_p$.  As $\xi$ approaches 0, the metric $d\bar s^2$
approaches
\be
d\bar s^2 = d\xi^2 + \fft{a^2}{4b^2}\, s^2\, (d\sigma + \cos\theta\, d\psi)^2
+ \ft14 (d\theta^2 + \sin^2\theta\, d\psi^2)\,.
\ee
In general, there will be a conical singularity at $\xi=0$, but this
is avoided if
\be
\fft{a}{b} = p\,.
\ee
As $\xi$ approaches the other endpoint, at $\xi=\ft12\pi$, the metric
$d\bar s^2$ approaches
\be
d\bar s^2 = d\xi^2 + \ft14 c^2\, \Big[ d\theta^2 + \sin^2\theta\,
d\psi^2 + (d\sigma+\cos\theta\, d\psi)^2\Big]\,.
\ee
This is locally $\R^4$, but there will be a conical singularity at
$\xi=0$ unless the lens space $S^3/Z_p$ is just $S^3$ itself; in other
words $p=1$.

    Thus we see that if $a=b$ and $p=1$, we have a completely
non-singular embedding of AdS$_5$ in eleven-dimensional supergravity.
In this case the warp factor $\wtd\Delta$ is simply equal to the
constant $a^2$.  For other choices of $p$ and the constants $a$ and
$b$, we have an M-theory embedding of AdS$_5$ that is ``almost''
non-singular, with relatively mild orbifold-like conical singularities
at $\xi=0$ and $\xi=\ft12\pi$.  In these more general cases the warp
factor $\wtd\Delta$ given in (\ref{ads5nonsing}) is a function of the
azimuthal coordinate $\xi$.  It is, however, always non-singular,
provided that $a\, b\ne0$.

   It is worth remarking that in the non-singular case $a=b$, $p=1$,
the metric $d\bar s^2$ is precisely the Fubini-Study metric on $CP^2$.
This and other geometrical aspects of the complex projective spaces
are discussed in the appendix.

   At the level of the AdS$_5\times S^5$ vacuum solution, the
untwisting of the fibres to give an AdS$_5\times CP^2\times S^1$
solution in the Hopf-duality related type IIA framework was already seen in
\cite{ads5s5}.  Here, we have gone further, and obtained the
Kaluza-Klein reduction Ansatz for the $SU(2)\times U(1)$ gauged
five-dimensional supergravity, viewed now as a $CP^2\times S^1$
reduction of type IIA supergravity.  As was discussed in
\cite{ads5s5}, there are some peculiarities associated with the type
IIA description, resulting from the fact that $CP^2$ does not admit an
ordinary spin structure.  This means that at the level of the
low-energy supergravities, there will be no fermions at all in the
type IIA description.  They will only be restored when the T-duality
is considered at the level of the full string theories, with the
fermions that carry charges with respect to the winding-mode vector
in the type IIB picture ending up in the Kaluza-Klein spectrum in the
type IIA picture.

   In fact this type of phenomenon is not confined to the fermionic
sector.  A bosonic example can be seen by looking at the reduction
Ansatz for the NS-NS and R-R 2-form potentials of the type IIB
theory.  In the $S^5$ reduction Ansatz in (\ref{10metans}), 
we saw that the 2-forms reduce as \cite{d5gauge}
\be
\hat A_\2^{NS} + \im\, \hat A_\2^{RR} = -s\, g^{-1} \, e^{\im\, g\,
\tau/\sqrt2}\, A_\2\,.\label{d52form}
\ee
Thus we
see that although the metric reduction Ansatz
(\ref{d5case1}) is invariant under both the $\del/\del z$ and
$\del/\del\tau$ $U(1)$, Killing symmetries, the 2-form Ansatz
(\ref{d52form}) is not invariant under the $\del/\del\tau$ symmetry.
Thus we would have to truncate $A_2$ from the five-dimensional theory
in order to carry out the previously-discussed  $S^1$ reduction.
In a similar fashion, one would find that the reduction Ans\"atze for
all the fermion fields would involve $\tau$-dependent complex
exponential factors, and thus would have to be truncated from the theory.

\subsection{Non-singular Hopf reduction of the $S^7$ embedding}

   We can also perform a non-singular Hopf reduction of the $S^7$
embedding of $N=4$ gauged $SO(4)$ four-dimensional supergravity.
Here, we may take linear combinations of the two
$S^1$ Hopf fibres in the two $S^3$ factors in the foliation of $S^7$,
so that for one of the two combinations the circle never degenerates to zero
radius, for any value of $\xi$.  This allows us to perform a
non-singular reduction of the $S^7$ Ansatz to give an Ansatz for the
non-singular embedding of $N=4$ gauged $SO(4)$ supergravity into type IIA 
supergravity.  As we shall see, in the limit where all the scalar and
gauged fields in $D=4$ are set to zero, this reduces to the
AdS$_4\times CP^3$ solution, which was discussed from a
string-theoretic viewpoint in \cite{susywosusy}.  A Kaluza-Klein
reduction at the level of linearised fluctuations around this background
was discussed in \cite{nilpop}.  Using our procedure here, we can
obtain the fully non-linear reduction Ansatz for the $N=4$ gauged
$SO(4)$ theory.

   Using the standard formulae
\bea
\sum_i h_i^2 &=& \sum_i (D\mu^i)^2 + 2 g^2\, (dz +
\cA_\1)^2\,,\nn\\
\sum_i \td h_i^2 &=& \sum_i (\wtd D\td\mu^i)^2 + 2 g^2\, 
(d\td z + \wtd \cA_\1)^2\,,
\eea
and then defining the linear combinations $z=x+y$, $\td z =-x+y$ as
the new $S^1$ coordinates, the eleven-dimensional metric Ansatz 
(\ref{metans}) becomes
\bea
d\hat s_{11}^2 &=& \Delta^{\fft23}\, ds_4^2 + 2 g^{-2}\,
\Delta^{\fft23}\, d\xi^2
+ \ft12
g^{-2}\, \Delta^{\fft23}\, \Big[ c^2\, \Omega^{-1}\, \sum_i (D\mu^i)^2
+ s^2\, \wtd\Omega^{-1}\, \sum_i (\wtd D\td\mu^i)^2\Big]\nn\\
&&+\Delta^{-4/3}\, \wtd\Delta \, \Big[dy + 
\fft{(c^2\, \wtd\Omega - s^2\, \Omega)\, dx + c^2\, \wtd\Omega\,
\cA_\1 + s^2\, \Omega\, \wtd\cA_\1}{\wtd\Delta}
\Big]^2\nn\\
&& +\Delta^{2/3}\, \wtd\Delta^{-1}\, s^2\, c^2
\, [(a^2+b^2)\, dx + a\, \cA_\1 -b\, \wtd\cA_\1]^2\,,
\label{metansnew}
\eea
where
\be
\wtd\Delta \equiv b^2\, c^2\, \wtd\Omega + a^2\, s^2\, \Omega\,.
\ee

    We can now perform the Hopf reduction on the $y$ coordinate.  This
gives
\bea
ds_{10}^2 &=& \Delta^{1/2}\, \wtd\Delta^{1/8}\,
\Big\{ ds_4^2 + 2 g^{-2}\, d\xi^2 + \ft12 g^{-2}\, c^2\, \Omega^{-1}\,
\sum_i (D\mu^i)^2 \nn\\
&&\qquad + \ft12 g^{-2}\, s^2\, \wtd\Omega^{-1}\, \sum_i
(\wtd D\td\mu^i)^2 + s^2\, c^2\, \wtd\Delta^{-1}\,
[(a^2+b^2)\, dx + a\, \cA_\1-b\, \wtd\cA_1]^2\Big\}\,,\nn\\
e^{\phi} &=& \Delta^{-1}\, \wtd\Delta^{3/4}\,.\label{ads42a}
\eea

   If we look at the vacuum solution where $X=1$, $\chi=0$,
$A_\1^{ij}=0$, $\wtd A_\1^{ij}=0$ and the metric $ds_4^2$ is taken to
be AdS$_4$, th type IIA metric in (\ref{ads42a}) takes the form
\be
ds_{10}^2 = \wtd\Delta^{1/8}\, ds_4^2 + 2g^{-2}\, \wtd\Delta^{1/8}\,
d\bar s^2 \,,
\ee
where
\bea
d\bar s^2 &=& d\xi^2 + \ft14 c^2\, (d\theta^2+\sin^2\theta\, d\psi^2) +
\ft14 s^2\, (d\td\theta^2 + \sin^2\td\theta\, d\wtd\psi^2) \nn\\
&&+ 
s^2\, c^2\, \wtd\Delta^{-1}\, [(a^2+b^2)\, dx + \fft{a}{\sqrt2 g}\,
\cos\theta\, d\psi - \fft{b}{\sqrt2 g}\, \cos\td\theta\,
d\wtd\psi]^2\,.
\eea
To avoid conical singularities on the level surfaces at constant $\xi$
we must require that the period $\delta x$ of the angular coordinate
$x$ should be such that
\be
\fft{\sqrt2\, g\, (a^2+b^2)}{a}\, \delta x = \fft{4\pi}{p}\,,\qquad
 \fft{\sqrt2\, g\, (a^2+b^2)}{b}\, \delta x = \fft{4\pi}{q}\,,
\ee
where $p$ and $q$ are integers, and so the ratio $a/b$ must be
rational:
\be
\fft{a}{b} = \fft{p}{q}\,.
\ee
Furthermore, if we wish to avoid conical singularities at the points
$\xi=0$ and $\xi=\ft12\pi$ where the level surfaces degenerate, we
must require that $p=q=1$.  (The discussion is analogous to that in
the previous subsection.)   Thus if $p=q=1$, implying that $a=b$, we
obtain a completely non-singular embedding of AdS$_4$ in the type IIA
theory.  In this case the warp factor $\wtd\Delta$ is simply equal to
the constant $a^2$.  For more general choices of $p$ and $q$, there
are mild orbifold-like conical singularities at $\xi=0$ and
$\xi=\ft12\pi$, and the warp factor $\wtd\Delta$ becomes dependent on
$\xi$.  (It is, however, non-singular, provided that $a$ and $b$ are
both non-zero.)

    Note that in the case where $p=q=1$ and $a=b$, the metric $d\bar
s^2$ is in fact the Fubini-Study metric on $CP^3$, written in a
particularly simple coordinate system.  This is discussed in more
detail in the appendix.   It should perhaps be emphasised that unlike
the $CP^2$ reduction of the previous subsection, here the $CP^3$
reduction does not lead to any loss of the (half-maximal)
supersymmetry.  This is related to the fact that $CP^3$, unlike
$CP^2$, does admit an ordinary spin structure.

    Although our principal purpose in this section was to examine
non-singular embeddings, we can also entertain the idea of making a
further $S^1$ reduction on the $x$ coordinate in the type IIA metric
(\ref{ads42a}).  This can be used in order to obtain another embedding
of the four-dimensional $N=4$ $SO(4)$-gauged supergravity in type IIB
supergravity.  Since the radius of the circle parameterised by $x$ 
vanishes both at $\xi=0$ and $\xi=\ft12 \pi$ this IIB embedding will
be a singular one.  In the string frame, the type IIB metric is
\bea
ds_{10}^2(IIB,str) &=& \wtd\Delta^{1/2}\,
\Big\{ ds_4^2 + 2 g^{-2}\, d\xi^2 + \ft12 g^{-2}\, c^2\, \Omega^{-1}\,
\sum_i (D\mu^i)^2 \nn\\
&&\qquad + \ft12 g^{-2}\, s^2\, \wtd\Omega^{-1}\, \sum_i
(\wtd D\td\mu^i)^2 + \fft{s^{-2}\, c^{-2}}{a^2+b^2}
\, dx^2\Big\}\,.
\eea

\section{Conclusions}

    In this paper, we obtained warped Kaluza-Klein embeddings of the
$D=7$, 6, 5 and 4 gauged supergravities with half-maximal
supersymmetry.  The characteristic feature of these warped embeddings
is that the vacuum solution where the lower-dimensional spacetime is
AdS has a non-trivial warp-factor multiplying the AdS metric, which
depends on one of the coordinates of the internal reduction manifold.
We constructed these warped embeddings by starting from the
previously-known spherical reductions that give rise to these
supergravities.  After performing circle reductions or T-duality
transformations on the Hopf fibres of $S^3$ submanifolds of the
original internal spheres, we obtained the new embeddings which can be
viewed as reductions on ``mirror manifolds'' dual to the original
spheres.  For all cases except $D=6$, the original Kaluza-Klein
reductions give non-warped solutions in the case of a pure AdS vacuum
solution.  Table 1 summarises the half-maximally supersymmetric
supergravities, their previously-known embeddings, and the new ones
that we obtained in this paper.

\bigskip\bigskip
\centerline{
\begin{tabular}{|c|c|c|c|c|}\hline
$D$ & 7 & 6 & 5 & 4 \\ \hline\hline
$N$ & 2 & 2 & 4 & 4  \\ \hline
$G$ & $SU(2)$  & $SU(2)$ & $SU(2)\times U(1)$ & $SO(4)$  \\ \hline\hline
Previously &   &    &    &     \\ 
Known  & M-theory & Massive IIA  & Type IIB & M-theory    \\ 
Embedding &       &              &     &              \\ \hline\hline
New       &       &              &     &              \\ 
Warped    & Type IIA   &Type  IIB & M-theory& Type IIB \\ 
Embedding &            &          &         &          \\ \hline
\end{tabular}}
\bigskip

\noindent{\bf Table 1:}
The half-maximally supersymmetric gauged supergravities in dimension
$D$, with $N$ supersymmetries and gauge group $G$.  Their previously
known Kaluza-Klein embeddings and the new warped embeddings obtained
in this paper are listed.
\bigskip\bigskip

   In the above warped embeddings, the warp factors can become
singular at the limits of the range of the internal azimuthal
coordinate $\xi$, since the Hopf fibres on which we performed $S^1$
reductions can approach zero radius at these endpoints.  However, we
also showed that in the cases $D=5$ and $D=4$, it is possible to
perform an $S^1$ Hopf reduction on a $U(1)$ fibre whose length remains
non-zero everywhere.  Thus in such cases the resulting warped
embedding can be non-singular.  The non-singular embeddings give rise,
in the case of a pure vacuum solution, to AdS$_5\times CP^2\times T^2$
in M-theory \cite{ads5s5} and AdS$_4\times CP^3$ in type IIA
supergravity \cite{nilpop,susywosusy}.  These solutions themselves are
in fact not warped.  We also obtained warped generalisations, at the
price of introducing rather mild orbifold-like singularities in the
internal manifolds.

    Recently, non-singular warped embeddings of AdS$_5$ in M-theory
\cite{fs,mn} and AdS$_3$ in type IIB supergravity \cite{mn} were
obtained.  The construction in \cite{mn} consists of finding an
AdS$_d\times \Sigma_g$ solution in $(d+2)$-dimensional gauged
supergravity, where $\Sigma_g$ is a Riemann surface, and then lifting
this to M-theory or type IIB using the Kaluza-Klein Ans\"atze
constructed in \cite{ten}.  These warped solutions are inequivalent to
any of the ones that we have obtained in this paper.  The construction
leads to solutions with non-singular warp factors, but its
applicability is restricted to AdS$_d$ with $d=3$ and 5.  Clearly it
cannot be applied to $d=7$, since there is no suitable gauged
supergravity in $D=9$.  The singularity of the warped embedding of
AdS$_7$ in type IIA that was discussed in this paper and in
\cite{clpv} may therefore be unavoidable.  In the case of AdS$_4$, the
analogous construction starting from AdS$_4\times \Sigma_g$ would
require a normal (unwarped) embedding of the six-dimensional $N=2$
$SU(2)$-gauged supergravity in a higher dimension.  However, as far as
is known, no such unwarped embedding exists \cite{d6gauge}.  The fact
that the only way to embed AdS$_6$ appears to be through a singular
warped configuration possibly suggests that singular warped embeddings
should not be overlooked.

\section*{Acknowledgements}

    We should like to thank Arta Sadrzadeh, Tuan Tran and Justin
V\'azquez-Poritz for useful discussions.  M.C. and C.N.P. are grateful
to Imperial College, London, H.L. and C.N.P. are grateful to CERN,
Geneva, and M.C. is grateful to CAMTP, Maribor, Slovenia, for support 
and hospitality.

\appendix

\section{$CP^{m+n+1}$ from $CP^m\times CP^n$}

     The metric on the unit $(p+q+1)$-sphere can 
be written in terms of a foliation of $S^p\times S^q$ for any $p$ and
$q$, as
\be
d\Omega_{p+q+1}^2 = d\xi^2 + c^2\, d\Omega_p^2 + s^2\, d\Omega_q^2\,,
\ee
where as usual $c\equiv \cos\xi$, $s=\sin\xi$, and the angle $\xi$
lies in the interval $0\le\xi\le \ft12 \pi$ (see, for example,
\cite{clpv}).  Furthermore, we know that if $p$ and $q$ are odd,
$p=2m+1$, $q=2n+1$, the metrics $d\Omega_p^2$ and $d\Omega_q^2$ on the
unit $S^p$ and $S^q$ spheres can each be written in terms of ``unit''
$CP^m$ and $CP^n$ Fubini-Study metrics\footnote{We define the ``unit''
Fubini-Study metric on $CP^n$ to be the one whose scale size is such
that its Hopf bundle gives the unit-radius $(2n+1)$-sphere.}
$d\Sigma_m^2$ and $d\Sigma_n^2$ as
\bea
d\Omega_p^2 &=& (d\tau_1+\cA_\1)^2 + d\Sigma_m^2\,,\nn\\
d\Omega_q^2 &=& (d\tau_2+\wtd\cA_\1)^2 + d\Sigma_n^2\,,\label{spherehopf}
\eea
where $d\cA_\1= 2J$ and $d\wtd\cA_\1=2\wtd J$, with $J$ and $\wtd J$
being the K\"ahler forms on $CP^m$ and $CP^n$.

    We can now follow the same procedure as in earlier sections, where
a $U(1)$ fibre coordinate is selected that is a linear combination of
those from the two spheres $S^p$ and $S^q$.  (We had $(p,q)=(3,1)$ for
the $S^5$ reduction of type IIB, and $(p,q)=(3,3)$ for the $S^7$
reduction of $D=11$ supergravity.)  To avoid unnecessary factors of 2,
we can just define $\tau_1=\ft12\psi +y$, $\tau_2=-\ft12\psi+y$ here, 
and hence we get
\be
d\Omega_{2m+2n+3}^2 = d\xi^2 + c^2\, d\Sigma_m^2 + s^2\, d\Sigma_n^2 +
s^2\, c^2 \, (d\psi+ \cA_\1- \wtd \cA_\1)^2 + (dy+B_\1)^2\,,
\ee
where
\be
B_\1\equiv \ft12 (c^2-s^2)\, d\psi + c^2\, \cA_\1 + s^2\, \wtd\cA_\1\,.
\ee
Since this metric on $S^{2m+2n+3}$ is now written as a $U(1)$ Hopf fibration 
(with unit radius for the fibres, whose coordinate is $y$), it follows
that the part of the metric orthogonal to $\del/\del y$ must be the
unit Fubini-Study metric on $CP^{m+n+1}$.  Thus we must have that
\be
d\Sigma_{m+n+1}^2 = d\xi^2 + c^2\, d\Sigma_m^2 + s^2\, d\Sigma_n^2 +
s^2\, c^2 \, (d\psi+ \cA_\1- \wtd \cA_\1)^2 \,.\label{cpmn}
\ee

    Since the construction of the unit sphere as the $U(1)$ Hopf fibration 
over a complex projective space goes as in (\ref{spherehopf}), it
furthermore follows that the K\"ahler form $\hat J$ on $CP^{m+n+1}$
will be given by
\be
\hat J \equiv  \ft12 d B_\1 = -s\, c\,d\xi\wedge(d\psi + \cA_\1 -\wtd\cA_\1) +
c^2\, J + s^2\, \wtd J\,.
\ee
Thus if we define the natural vielbeins $\hat e^A$ for the metric
$d\Sigma_{m+n+1}^2$ in (\ref{cpmn}), namely
\be
\hat e^0 = d\xi\,,\qquad \hat e^1 = s\, c\, (d\psi+ \cA_\1 - \wtd
\cA_\1)\,,\qquad \hat e^a = c\, e^a\,,\qquad \hat e^{\td a} = s\,
e^{\td a}\,,
\ee
where $e^a$ and $e^{\td a}$ are vielbeins for $d\Sigma_m^2$ and
$d\Sigma_n^2$, then we have
\be
\hat J = -\hat e^0\wedge \hat e^1 + c^2\, J + s^2\, \wtd J\,.
\ee
In other words, the non-vanishing vielbein components $\hat J_{AB}$ of the 
K\"ahler form on $CP^{m+n+1}$ are given by
\be
\hat J_{01} =-1\,,\qquad \hat J_{ab} = J_{ab}\,,\qquad \hat J_{ij} =
\wtd J_{ij}\,.
\ee
A simple calculation shows that the curvature 2-form for the metric
(\ref{cpmn}) on $CP^{m+n+1}$ indeed has the form,
\be
\hat \Theta_{AB} = \hat e^A\wedge \hat e^B + \hat J_{AC}\, \hat
J_{BD}\, \hat e^C\wedge \hat e^D + \hat J_{AB}\, \hat J_{CD}\, \hat
e^C\wedge \hat e^D\,,
\ee
which one expects for the unit Fubini-Study metric.  In terms of
vielbein components, we see that the Riemann tensor has the
characteristic structure for a space of constant holomorphic sectional
curvature \cite{kn},
\be
\hat R_{ABCD} = \delta_{AC}\, \delta_{BD} -\delta_{AD}\, \delta_{BC} +
\hat J_{AC}\, \hat J_{BD} -\hat J_{AD}\, \hat J_{BC} + 2\hat J_{AB}\,
\hat J_{CD}\,.
\ee

   This study of Fubini-Study metrics encompasses various
previously-known results, as well as providing many new ones.  For
example, if we take $m=0$, $n=1$, we get the metric on $CP^2$ written
(after sending $\psi\longrightarrow \psi/2$ for convenience)
as\footnote{$CP^1$ is the same as $S^2$.  Note, however, that the {\it
unit} $CP^1$, which we have defined to be such that its Hopf bundle
gives the unit-radius 3-sphere, is consequently a 2-sphere of radius
$\ft12$, whose metric is $d\Sigma_1^2 = \ft14(d\theta^2 + \sin^2\theta\,
d\varphi^2)$.}
\be
d\Sigma_2^2 = d\xi^2 + \ft14 s^2\, (d\theta^2 + \sin^2\theta\,
d\varphi^2) + \ft14 s^2\, c^2\, (d\psi + \cos\theta\,
d\varphi)^2\,,
\ee
which was first obtained in this form in \cite{gibpop} (with the
coordinate $r$ of that paper related to $\xi$ by $r=\tan\xi$).  The
general class of cases $m=0$, with $n$ arbitrary, was obtained in
\cite{orphans}; it gives an iterative expression for the Fubini-Study
metric on $CP^{n+1}$ in terms of that on $CP^n$.

   As a new example, we may obtain the following expression for the
Fubini-Study metric on $CP^3$, by taking $m=n=1$ (and sending
$\psi\longrightarrow \psi/2$ for convenience):
\bea
d\Sigma_3^2 &=& d\xi^2 + \ft14 c^2\, (d\theta^2 + \sin^2\theta\,
d\varphi^2) + \ft14 s^2\, (d\td\theta^2 + \sin^2\td\theta\,
d\td\varphi^2)\nn\\
&& +\ft14 s^2\, c^2\, (d\psi - \cos\theta\, d\varphi +
\cos\td\theta\, d\td\varphi)^2\,.\label{cp3met}
\eea
The K\"ahler form is given by
\be
\hat J = -\ft12 s\, c\, d\xi\wedge (d\psi - \cos\theta\,d\varphi +
\cos\td\theta\, d\td\varphi) + \ft14 c^2\, \sin\theta\, d\theta\wedge
d\varphi + \ft14 s^2\, \sin\td\theta\, d\td\theta\wedge d\td\varphi\,.
\ee

    The metric (\ref{cp3met}) reveals some interesting features of the
geometry of $CP^3$.  At each end of the $\xi$ coordinate range,
$0\le\xi\le\ft12 \pi$, the metric approaches a product of a
smooth $\R^4\times S^2$; for example at $\xi\approx 0$ we have
\be
ds^2 \approx d\xi^2 + \sin^2\xi\, d{\Omega_3'}^2 + \ft14
d\Omega_2^2\,,
\ee
with
\be
d{\Omega_3'}^2 = \ft14 \Big[d\td\theta^2 + \sin^2\td\theta\, d\td\varphi^2 
+ (d\psi + \cos\td\theta\, d\td\varphi -\cos\theta\,
d\varphi)^2\Big]\,.
\ee
Thus the terms $d\xi^2 + \sin^2\xi \, d{\Omega_3'}^2$ approach $\R^4$
as $\xi$ tends to zero, described in hyperspherical polar
coordinates.  There is a ``twist'' in the $U(1)$ fibres of the $S^3$
metric $d{\Omega_3'}^2$, involving the topologically non-trivial
Dirac monopole bundle over the $S^2$ factor $d\Omega_2^2 = d\theta^2 +
\sin^2\theta\, d\varphi^2$.  An analogous phenomenon occurs at the
other endpoint, at $\xi=\ft12\pi$.  This form of the $CP^3$ metric
is precisely the one that arises in the non-singular embedding of the
four-dimensional $N=4$ $SO(4)$-gauged supergravity in section 6.2.

    Another interesting aspect of the geometry of $CP^3$ that can be
seen from (\ref{cp3met}) is that each foliating surface at constant
$\xi$ has the structure of the manifold $Q(1,1)$ (sometimes known as
$T^{11})$, which is defined as the $U(1)$ bundle over $S^2\times S^2$
where the fibres have winding number 1 with respect to both of the
$S^2$ factors in the base.  The metric
\be
ds_5^2 = \fft1{\Lambda}\, (d\theta^2 + \sin^2\theta\, d\varphi^2) +
\fft1{\wtd\Lambda}\, (d\td\theta^2 + \sin^2\td\theta\, d\td\varphi^2) 
+ c^2\, (d\psi -\cos\theta\, d\varphi +\cos\td\theta\, d\td\varphi)^2 
\ee
on this manifold is homogeneous for any choice of the constants
$\Lambda$\, $\wtd\Lambda$ and $c$, and it is Einstein if
$\Lambda=\wtd\Lambda= 2/(3c^2)$ (see, for example, \cite{ads5s5}).
Thus as the coordinate $\xi$ ranges over the interval $0<\xi<\ft12\pi$
in (\ref{cp3met}), the foliating surfaces correspond to $Q(1,1)$ with
varying non-singular homogeneous ``squashings.''  None of the
foliating surfaces corresponds to the Einstein metric on $Q(1,1)$.

   Note that in general, the level surfaces at constant $\xi$ in the
metric (\ref{cpmn}) are the higher-dimensional generalisations of the
$Q(1,1)$ space, namely $U(1)$ bundles over $CP^m\times CP^n$, with
winding number 1 over each factor.

\end{document}